\documentstyle[11pt,newpasp,twoside,psfig]{article}

\index{Hammer, F.}

\begin{document}

\title{Morphological and Star Formation Evolution to z=1}
\author{F. Hammer}
\affil{DAEC, Observatoire de Paris-Meudon}

\begin{abstract}
The decrease, since z=1, of the rest-frame UV luminosity density is
related to global changes in morphology, color and emission lines properties
of galaxies. This is apparently followed by a similar decrease of the rest-frame
IR luminosity density. 

I discuss the relative contribution from the different galaxy morphological 
types to the observed evolution. The main contributors are compact 
galaxies observed in large number at optical wavelengths, and the sparse population
of extincted \& powerful starbursts observed by ISO. This latter population
is made of large and massive galaxies mostly found in interacting systems, 
some of which could be leading to the formation of massive ellipticals at z $<$ 1.

\end{abstract}


\section{Introduction}

The evolution of luminosity densities has been examined by Lilly et al (1996),
from 600 I$<$22 galaxies of the Canada France Redshift Survey (CFRS).
They found a large decrease by factor 10 of the rest-frame UV luminosity
density from z=1 to z=0. This factor has probably to be lowered to $\sim$6, since
 recent estimates (Treyer et al, 1998) of the local UV luminosity density are 
 1.5 times larger than previous estimates based on H$\alpha$ luminosity density 
 (Gallego et al, 1995).\\
 At 15$\mu$m deep counts show a steep slope below 400$\mu$Jy (see Elbaz et al, 
 1999). Associated with the flattening of the deep radio count slope (Fomalont et al,
 1991), this suggests the presence of an evolving population at infra-red wavelengths.
 On the basis of a sample of $\sim$ 30 15$\mu$m and radio sources, Flores et al
 (1999) have shown that the rest-frame IR luminosity density evolves as rapidly as
 the rest-frame UV luminosity density.\\
 It is important to notice that all these works are based on the observations of 
 relatively luminous galaxies in optical ($M_{B}<$ -20), as well as in IR ($L_{bol}>$
 2 $10^{11}$ $L_{\odot}$). And that the corresponding evaluations of
 luminosity density evolution are, strictly speaking, related to luminous galaxies.
 Assuming an unevolved shape of the luminosity function in both UV and IR
 would provide an equipartition of the energy output (or star formation rate density)
 between UV and IR light from z=1 ($\sim$ 9 Gyr ago) to the present day (Hammer,
 1999), in accordance with bolometric measurements of the background (see Pozetti
 et al, 1998).     

\section{Galaxy morphologies and their global evolution}

\subsection{Observations of the Hubble Space Telescope and their limitations}

Studies of distant galaxies are limited by the spatial resolution, since 1 pixel
 of HST/WFPC2 corresponds to 1$h_{50}^{-1}$kpc at z$\ge$0.75. This
provides the most severe limitation to their morphological studies. For
example, at z=0.75, a 5 kpc half-light radius would correspond to only 5 WFPC2 pixels,
with an HWHM of only two pixels. This limits the accuracy of bulge/disk deconvolution
for a non-negligeable fraction of the distant galaxies.\\
Another limitation is related to redshift dependent effects: for example at z=0.9 the
I (F814W) filter samples the rest-frame B band, a color which is more sensitive to
star forming regions. 24\% of the spirals observed at z$\sim$ 0.9 could
be mis-identified as irregulars (Brinchman et al., 1998) when compared to lower z 
systems.\\
Other effects (biases against low surface brightness objects or extincted disks)
  are caused by the limited photometrical depth reachable in a reasonable
  exposure time (few hours) and all the above limitations
emphasize the need for an optical camera optimized at the diffraction 
limit on an 8 meter space telescope.  

\subsection{Evolution of averaged properties}

Brinchman et al (1998) have presented the HST imagery of $\sim$ 340 I$\le$22
galaxies, spanning a redshift range from z=0.1 to z=1. Galaxy morphologies have
been classified by eye as well as through bulge/disk deconvolution. 
Brinchman et al (1998) quoted that 9\% of galaxies
at 0.2$<$ z $<$0.5 are irregulars, a fraction which reaches
32\% at 0.75$<$ z $<$1.
The luminous galaxies in the highest redshift bins were much bluer and with
a later type than that of a Sbc, conversely to present-day galaxies (Figure 1). Present-day
stellar population has an average color ($(B-K)_{AB}$=2.5) typical of an Sab (see 
Hammer, 1999).

\begin{figure*}[!h]
\centerline{\psfig{file=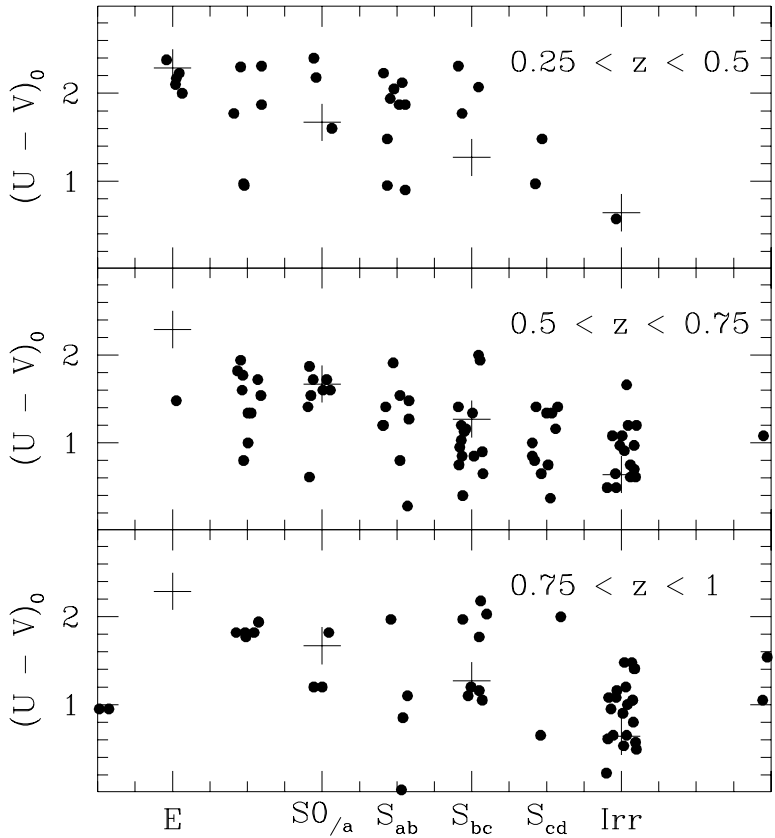,width=4in}}
\caption{ Rest frame $(U - V)_{AB}$ color versus HST morphological
class for $M_{B}\le$ -20 CFRS, in three redshift bins. Brinchman et al
classification is supported by the comparison with local values (large
crosses; Coleman et al, 1980). Most galaxies have an earlier type than Sbc 
in the lower redshift bin, while the reverse is found at high redshift. }
\end{figure*}
  
This trend is followed by a large redshift increase of the rate of emission line galaxies
(those with $W_{0}(OII)>$ 15\AA~) from 13\% locally to more than 50\% at z$>$0.5
 (Hammer et al, 1997). These properties, taken together, are consistent with
 the observed rest-frame UV luminosity density, and confirm a declining star
 formation history since the last 9 Gyr.

\section{Evolution of galaxies selected by morphology}

\subsection{Ellipticals}

The number density evolution of luminous ellipticals is still controversial (see
Kauffman et al, 1998). It is an important debate, because the monolithic
collapse scenario (see e.g. Bower et al., 1992) predicts their formation at a 
high redshift conversely
to hierarchical models (see White and Rees, 1978) in which massive ellipticals
are formed at later times from the collapse of smaller units. The two scenarios
are to predict a different star formation history, since a large fraction of
the metals are bound in bulges (see Fukugita et al, 1998).\\
In selecting elliptical galaxies on the basis of their luminosity profiles, Schade
et al. (1999) have shown that a color criterion is rather unefficient. This latter 
likely selects as many disks (possibly with
small amounts of dust) as ellipticals. Schade et al. also find no evidence for a decline
in space density of ellipticals since z=1, although this conclusion is limited to
the small sample of objects in consideration (46 galaxies). More interesting
is the fact that a third of the selected ellipticals show significant 
emission lines, which they interpret as related to small events of star formation
  at z$<$1, representing the formation of only few percents of the stellar mass.\\
It is premature to conclude on their number density evolution before a larger sample is
gathered. Several biases can also affect the apparent density of ellipticals at
high z, including a possible mis-classification of some S0 with faint disks.
Even the detection of small amounts of star formation in z$<$1 ellipticals 
should be taken with caution,
because the presence of emission lines seems not to affect their (U -V) colors
(Figure 2). Extinction of hot stars might be at work in these objects, but
this should be rather complex to explain the presence of the [OII]3727 emission line. 
Alternatively, these emission lines could be as well related to the presence of
an AGN, which is suggested to be present in most of the massive ellipticals by
Hammer et al (1995), on the basis of their radio inverted spectra. There is
however an evidence that elliptical galaxies are not contributing to the
observed evolution of the rest-frame UV luminosity density.

\begin{figure*}[!ht]
\centerline{\psfig{file=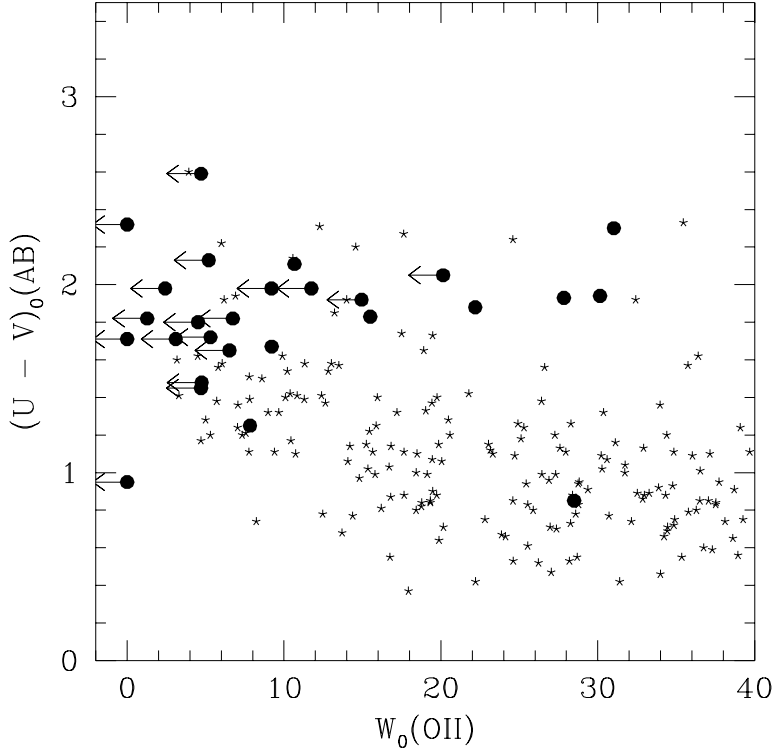,width=4in}}
\caption{ Rest frame $(U - V)_{AB}$ color versus OII rest frame equivalent
widths for $M_{B}\le$ -20 from z=0.5 to z=1. Full dots represents ellipticals
selected by Schade et al (1999), and small crosses are other CFRS galaxies
observed with the HST. Arrows indicate a limit in the OII equivalent width. In
 a star forming event, the
correlation of the OII EW with (U-V) color for the CFRS galaxy population is
expected. No such a correlation is seen in the
selected ellipticals. }
\end{figure*}    

\subsection{Large disks}

The density of large disks
 with $r_{disk}\ge$ 3.2$h_{50}^{-1}$kpc is found to be the same at z=0.75 as
locally (Lilly et al, 1998). Only a density decrease by less than 30\% 
at z=1 is consistent with the data. Lilly et al (1998) also find that 
UV luminosity density produced by large disks shows only a modest increase with the redshift.
There is however a general shift towards later type for disks in
the highest redshift bin.\\
 From long-slit spectroscopy studies, Vogt et al (1997,
see also Koo, 1999) show an unevolved Tully Fischer
relation for disks at z $\sim$ 1. However the disk velocity could be affected
by the presence of companions at high z as well as by the geometry and
alignement of the slit with the disk major axis (see Amram et al, 1996). Higher
 resolution spectroscopy associated with integral field unit will definitively
 establish the Tully Fisher relation at high redshift.  \\

An important question is to know if the present-day population of galaxies, similar
 to the Milky Way, was already in place 7 to 9 Gyr ago. Studies of
the Milky Way (see Boissier and Prantzos, 1999) as well as the Schmitt law for
disks (see Kennicutt, 1998) argue in favor of a rather long duration (3-7 Gyr) for
the formation of the bulk of their stars. Number density evolution, Tully Fisher relation 
and present-day properties of disks, they all suggest a relatively passive evolution
of large disks. Redshift changes in large disks 
appear not to be the main contributors to the evolution of the rest-frame
UV luminosity density, as detected in that redshift range. However, it is
still unclear if all the disks observed at z=1 are progenitors of present-day
disks. Star formation estimates in disks galaxies might be severely 
affected by extinction, as shown by Gruel et al (1999, in preparation).\\

\section{The major contributors to the observed evolution}

\subsection{Compact galaxies}

\begin{figure*}[!h]
\centerline{\psfig{file=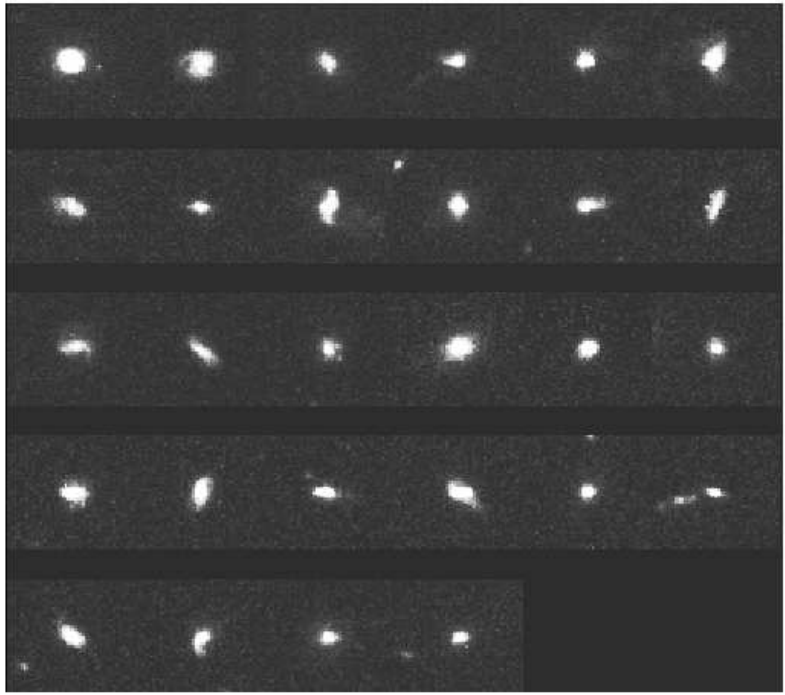,width=4.5in}}
\caption{ 28 compact galaxies with z$>$0.4, $M_{B}<$-20.5 and
color bluer than an Sbc. Each stamp has a size of 40 $h_{50}^{-1}$kpc, and
galaxies are shown by increasing redshift (z=0.4 in the top left, z=1 in
the bottom right).}
\end{figure*}    

The most rapidly evolving population of galaxies detected in the visible is made of
small and compact galaxies with half light radius smaller than 5$h_{50}^{-1}$kpc
(Guzman et al, 1997; Lilly et al, 1998). Their UV
 luminosity density was 10 times higher at z=0.875 than at
z=0.375, and they correspond to $\sim$ 40\% of the rest-frame UV luminosity
density in the higher redshift bin (Hammer and Flores, 1998). These objects 
are somewhat enigmatic: their sizes -$r_{disk}\le$ 2.5$h_{50}^{-1}$kpc- and
their velocity widths -35 to 150 km/s (Phillips et al, 1997)- are apparently similar 
to those of local dwarves, while they are 10 to 100 times more luminous 
than a $M_{B}$=-17.5 dwarf.\\
Guzman (1999) has argued that compact galaxies are the result of bursts in low
 massive systems (few $10^{9}$$M_{\odot}$), which would generate
the present day population of spheroidal dwarf galaxies. Kobulnicky and
Zaritsky (1999) have estimated a range of Z=0.3$Z_{\odot}$ to $Z_{\odot}$ for
the metal abundance of  few z$<$0.5 compact galaxies. These values
 -as well as their luminosities- are rather consistent with those of
 local spiral galaxies or of the most massive irregular galaxies. An important question 
is to know if the very narrow emission lines are indeed sampling the gravitational
potential, or if alternatively they are only located in a small area of the galaxies,
or being affected by dust or inclination effects. An important fraction (if not all) of the
luminous compact galaxies at z$>$0.5 show evidences for low surface brightness
extents (Figure 3), as well as for a noticeable fraction of galaxies with companions. Further
studies with large exposure time at 8 meter telescope are required to study 
their continuum properties (absorption lines).

\subsection{Interacting and starbursting galaxies detected by ISO}

During a follow-up study with ISOCAM of the CFRS, Flores et al (1999) have detected
 galaxies with strong emission at
both radio and mid-IR wavelengths. They interpret them as being strong star forming
galaxies with SFR from 40 to 250 $M_{\odot}$$yr^{-1}$, most of their UV light
being reprocessed by dust to IR wavelengths. These galaxies represents 4\% of the
luminous ($M_{B}<$-20) galaxy population, while they produce as many stars as 
the rest of galaxy population.\\
Most of these star-forming galaxies 0.5$\le$z$\le$ 1,
appear to be strong mergers, or at least they show signs of interactions 
(Figure 4). It is important to notice that individual galaxies involved
in these systems have sizes larger than the normal galaxy population (Figure 5).
This argues in favor of the formation at z$<$1 of large systems, including
massive ellipticals by merging of two large disks. Several large disks are
also strong IR emitters, implying that UV luminosity samples only a small
fraction of their star formation.
 
\begin{figure*}[!ht]
\centerline{\psfig{file=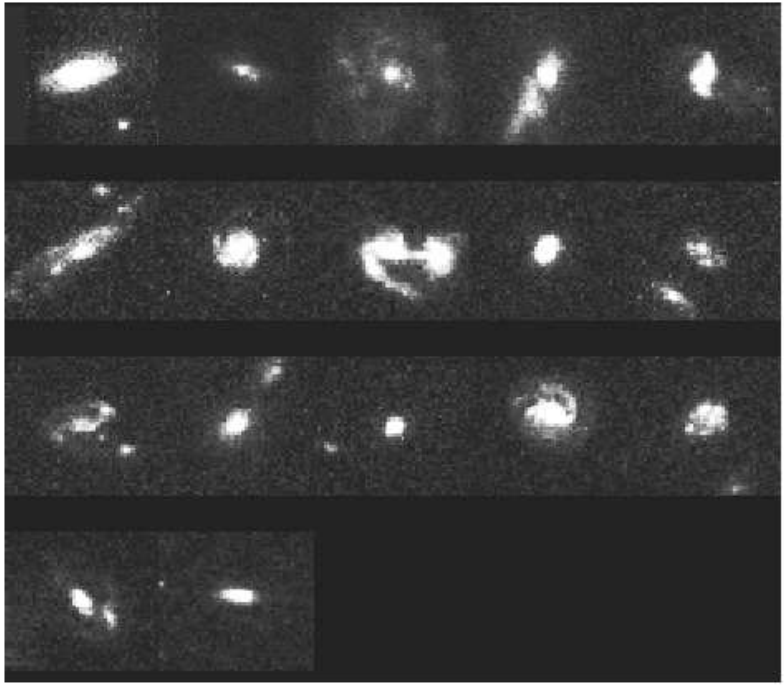,width=4.5in}}
\caption{ 17 starburst galaxies with $f_{15\mu}\ge$350$\mu$Jy and/or
$f_{5GHz}\ge$16$\mu$Jy from Flores et al (1999) and Flores and Hammer (2000,
in preparation). Each stamp has a size of 40 $h_{50}^{-1}$kpc, and
galaxies are shown by increasing redshift (z=0.4 in the top left, z=1 in
the bottom right).}
\end{figure*}    

\begin{figure*}[!ht]
\centerline{\psfig{file=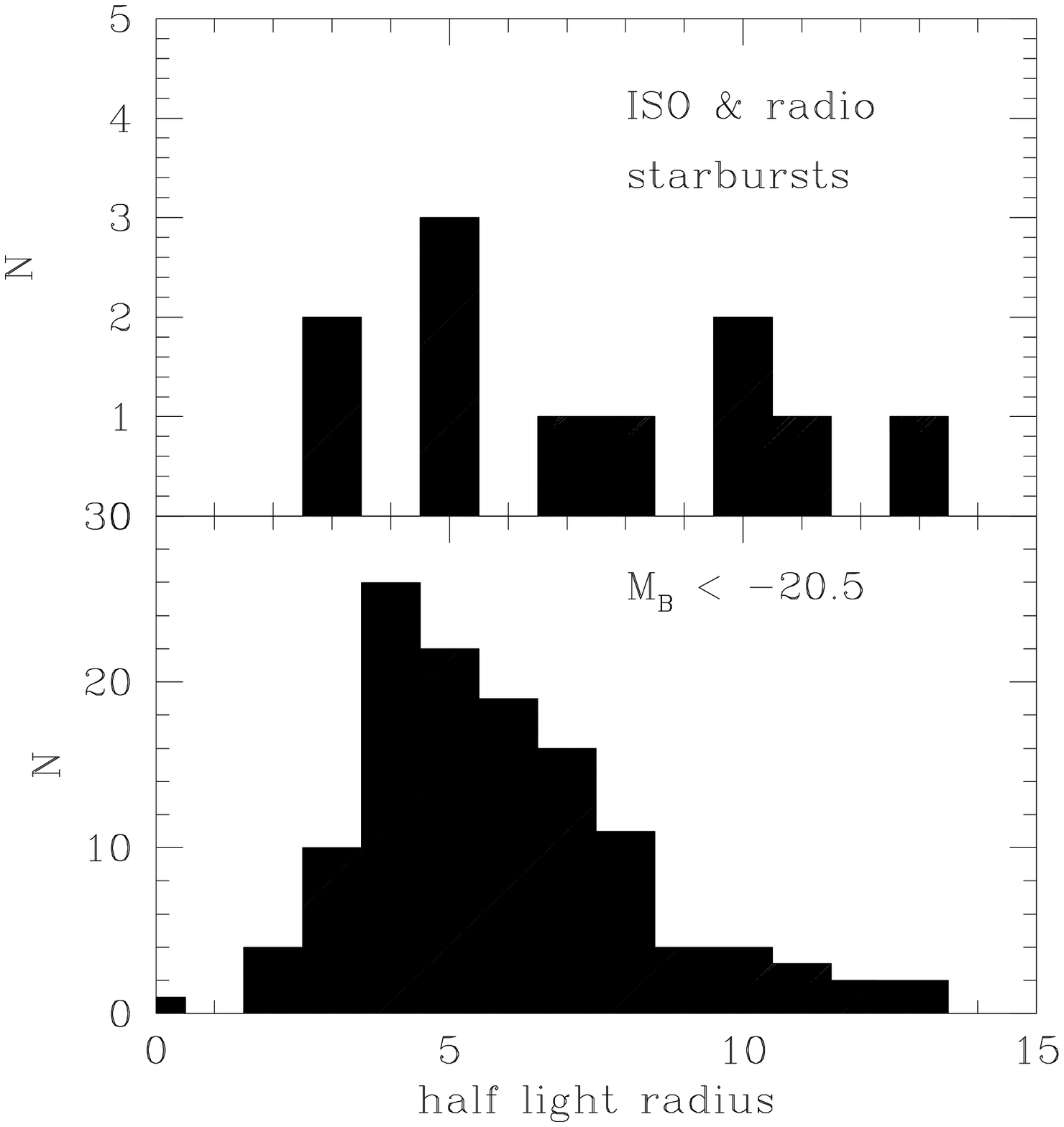,width=4in}}
\caption{ ({\it top}) Half light radius distribution for starburst galaxies detected
by ISO and VLA. Only the major component in the interacting system is taken into account. 
({\it bottom}) Same distribution for all CFRS galaxies with $M_{B}\le$-20.5 and
0.4$<$z$<$1.}
\end{figure*}
    
\section{Conclusion}

UV and IR luminosity density both present a surprisingly similar evolution since
the last 9 Gyr. The former is dominated by a numerous population of blue galaxies, the
latter is concentrated in a small fraction of the galaxy population, mostly interacting and
dusty galaxies.\\
When looking at the morphological properties of the population responsible for the
luminosity density evolution, UV and IR selected galaxies draw strikingly different
pictures:\\
-Large galaxies -elliptical and large disks- have blue or UV properties almost unchanged
since z=1, and most of the reported evolution in the UV is related to irregular and
compact galaxies.\\
-Conversely, most of the galaxies responsible for the IR luminosity density evolution,
are large galaxies (from S0 to Sbc), generally found in interacting systems; they
include some good candidates to the formation of a massive elliptical at z $<$1 resulting
from the collapse of two disk galaxies.\\
It is too premature to test which scenario -monolithic collapse or hierarchical
 model- is dominating the galaxy formation. But there is good evidence that at least,
 galaxy interactions were still at work in the latest 9 Gyr to form massive galaxies.
 Larger samples and better spectroscopic resolution are required to quantify
 the above observational facts.

\acknowledgments

I would like to thank the organizing and scientific committees for their
kind invitation.

\end{document}